\PassOptionsToClass{balance=false}{acmart}
\documentclass[sigconf]{acmart}
\acmYear{2027}
\copyrightyear{2027}
\usepackage[utf8]{inputenc}
\usepackage{multirow}
\usepackage{graphicx} 
\usepackage{booktabs}
\usepackage{amsmath}
\usepackage{enumitem}
\usepackage{array}
\usepackage{colortbl}
\usepackage{pifont}
\definecolor{checkgreen}{RGB}{21,128,61}
\definecolor{crossred}{RGB}{190,35,45}
\definecolor{oursbg}{RGB}{230,244,233}
\definecolor{oursgoodbg}{RGB}{198,232,205}
\newcommand{\yesmark}{\textcolor{checkgreen}{\ding{51}}}
\newcommand{\nomark}{\textcolor{crossred}{\ding{55}}}
\providecommand{\ours}{}
\renewcommand{\ours}{\textsc{CIVER}}
\usepackage{url}
\graphicspath{{figures/}}
\title{\ours{}: Causal Intervention Verification for Reliable Code Review}
\author{Xinyuan Song}
\affiliation{\institution{Causal Dynamics Lab}\country{USA}}

\author{Hasibul Haque}
\affiliation{\institution{Causal Dynamics Lab}\country{USA}}

\author{Jay Somasundaram}
\affiliation{\institution{Causal Dynamics Lab}\country{USA}}

\author{Ryan Turner}
\affiliation{\institution{Causal Dynamics Lab}\country{USA}}

\author{Bowen Zhu}
\affiliation{\institution{Causal Dynamics Lab}\country{USA}}

\author{Liang Zhao}
\affiliation{\institution{Causal Dynamics Lab}\country{USA}}

\setcopyright{none}
\acmConference[WWW]{The ACM Web Conference}{Manuscript}{Under review}
\acmDOI{}
\acmISBN{}
\AtEndPreamble{\hypersetup{
  colorlinks=true,
  linkcolor=blue,
  urlcolor=blue,
  filecolor=blue,
  citecolor=black,
  citebordercolor={0 1 0},
  pdfborder={0 0 0},
  pdfborderstyle={}
}}
\makeatletter
\AtBeginDocument{%
  \pretocmd{\hyper@natlinkstart}{\begingroup\def\@pdfborder{0 0 1}}{}{}%
  \apptocmd{\hyper@natlinkend}{\endgroup}{}{}%
}
\makeatother
\begin{document}
\begin{abstract}
Code review is essential for maintaining software repositories by identifying defects, improving code quality, and supporting reliable software evolution. With the growing adoption of LLM-based coding agents, an increasing number of code contributions and repository changes are generated by these agents. Consequently, LLM-based code review agents are becoming increasingly prevalent to evaluate these contributions and maintain software quality. However, LLM reviewers often generate lengthy reviews containing numerous potential issues, many of which are false positives or lack sufficient evidence. These unreliable reviews can mislead developers and downstream coding agents, motivating the need to distinguish actionable findings from unsupported claims. We propose \ours{}, a causal intervention-based framework that identifies the core issues underlying code-review findings and distinguishes genuine defects from false positives. \ours{} integrates three modules: a review module assesses findings, an intervention module generates repair-oriented perturbations, and a verifier module scores review reliability. Through joint optimization via expert iteration, the three modules identify genuine defects by responding to targeted repairs while remaining stable under unrelated changes. Extensive experiments on SWE-Review-Bench demonstrate that \ours{} correctly identifies 77\% of genuine defects through targeted repairs. Cross-project and downstream repair evaluations further demonstrate that \ours{} generalizes across repositories and improves coding agents' task success rates, highlighting its effectiveness for reliable automated code review.
\end{abstract}

\maketitle

\section{Introduction}
\label{sec:intro}

\begin{figure*}[t]
\centering
\includegraphics[width=0.8\textwidth]{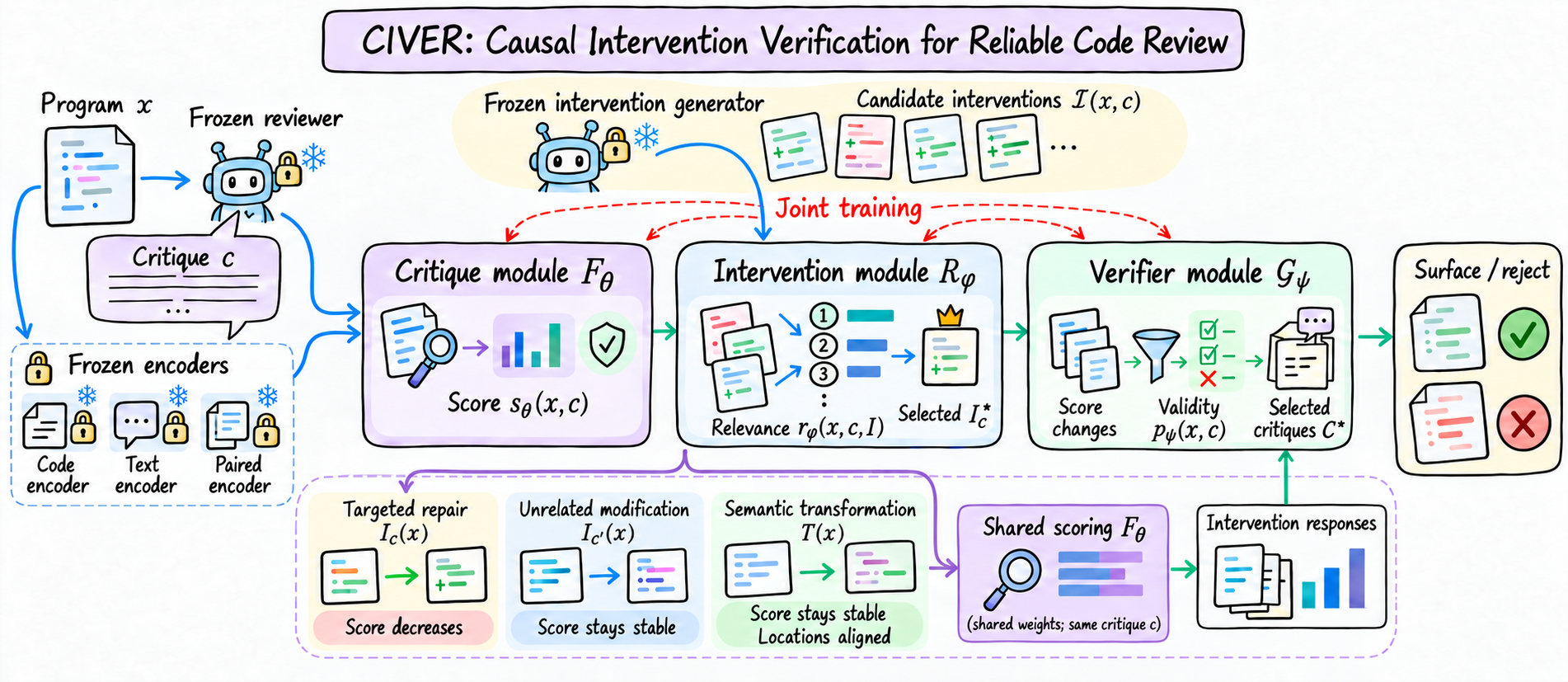}

\caption{\textbf{Overview of \ours{}.} (a) A frozen LLM reviewer generates code reviews, which are transformed into candidate critiques by the review module. (b) The intervention module selects relevant code modifications, including targeted repairs and unrelated perturbations. (c) The verifier module evaluates review reliability by comparing scores across original and modified program states. The three modules are jointly optimized through expert iteration to distinguish genuine defects from false positives, while pretrained models remain frozen.}
\Description{A three-module code-review pipeline with a review module F-theta, an intervention module R-phi and a verifier module G-psi. Frozen components supply critiques, encodings and edit proposals. Shared critique scoring compares original, target-repaired, unrelated-repaired and mapped transformed states; the verifier aggregates the responses and selects final findings. Backward learning paths connect all three modules.}
\label{fig:pipeline}

\end{figure*}

Code review is a fundamental component of collaborative software development~\citep{bacchelli2013expectations,rigby2013convergent}, enabling developers to discuss proposed changes, identify defects, and maintain software quality through pull requests (PRs) on Web-based platforms such as GitHub. Recent advances in large language model (LLM)-based coding agents, including SWE-agent~\citep{yang2024sweagent}, OpenHands~\citep{wang2024opendevin}, Claude Code~\citep{anthropic2025claudecode}, and Codex~\citep{openai2025codex}, have enabled increasingly autonomous software development, from implementing features and fixing bugs to maintaining entire repositories. As these agents become widely adopted, they generate a growing volume of code contributions, including commits, PRs, tests, and documentation updates. Recent empirical studies~\citep{robbes2026agentic,watanabe2025agentic} report increasing adoption of coding agents across GitHub repositories and their participation in real-world software contributions. For example, GitHub's Octoverse 2025 report~\citep{github2025octoverse} documents over one million PRs created by its Copilot coding agent within five months of its introduction. This growing volume of agent-generated contributions creates an increasing demand for scalable code review, driving the adoption of LLM-based review agents such as OpenCodeReview~\citep{li2026opencodereview},
GitHub Copilot Code Review~\citep{github2025copilotreview},
and CodeRabbit~\citep{cihan2024automated} to evaluate code changes and support reliable repository maintenance. Recent benchmarks, including SWE-Review-Bench~\citep{wang2026swereview}, AACR-Bench~\citep{zhang2026aacrbench}, and CodeReviewBench~\citep{kodus2026codereviewbench}, further highlight the importance of evaluating automated code review in realistic software engineering workflows~\citep{khan2026survey}.

Despite their growing capabilities, LLM-based reviewers~\citep{cihan2025evaluating} frequently generate lengthy reviews containing numerous potential issues, many of which are false positives~\citep{jin2026overcorrection,mao2026youreright}, redundant observations, or claims unsupported by the actual program behavior. Unlike traditional code analysis, where findings can often be associated with explicit program constraints, LLM-generated critiques are expressed in natural language and may appear convincing even when the underlying defect does not exist. Recent empirical studies on LLM-assisted code review~\citep{adalsteinsson2025rethinking} and noisy review comments~\citep{liu2025toonoisy} further highlight the prevalence of non-actionable feedback and concerns about review reliability. This creates a critical challenge for automated code review: generating more critiques does not necessarily improve review quality. Instead, excessive unreliable findings increase the burden of distinguishing genuine defects from unsupported concerns. More importantly, when these critiques are passed to downstream coding agents, false positives~\citep{charoenwet2026agentic,jin2026overcorrection} may trigger unnecessary modifications, introduce new defects, or disrupt previously correct functionality. Therefore, reliable automated code review requires not only identifying potential issues but also verifying which critiques correspond to genuine defects and should guide subsequent code modifications.

Existing approaches have explored several directions for improving code review reliability. LLM-based review systems, such as CodeReviewer~\citep{li2022codereviewer}, SGCR~\citep{wang2025sgcr}, and OpenCodeReview~\citep{li2026opencodereview}, leverage code representations and repository context to generate more informative critiques. Self-Refine~\citep{madaan2023selfrefine} improves model outputs through iterative feedback, while Refute-or-Promote~\citep{agarwal2026refuteorpromote} and adversarial review~\citep{qiu2026adversarialreview} employ independent or adversarial verification to assess candidate findings. Other approaches, including LLM4PFA~\citep{du2025llm4pfa}, execution-path-based false-positive reduction~\citep{du2026reducingfp}, and fix-guided verification~\citep{jin2026overcorrection}, investigate program feasibility and proposed repairs to filter unreliable findings. Counterfactual code analysis~\citep{egea2025vision} and ComPass~\citep{zhang2026compass} further explore program-state comparisons and semantic consistency for vulnerability detection and patch assessment. However, these methods do not explicitly establish whether a critique is causally connected to its claimed defect. High reviewer confidence or agreement among verification agents may reinforce incorrect assumptions, while changes in critique scores after code modifications may reflect superficial differences rather than the removal of the underlying defect.

To address this limitation, we investigate code-review reliability from a causal intervention perspective. A reliable critique should respond to repairs targeting its claimed defect while remaining stable under unrelated changes. For example, if a critique identifies a missing null check, introducing the necessary guard should invalidate the original finding, whereas modifying an unrelated function should not affect its validity. Conversely, if the guard already exists, the critique should be rejected rather than accepted simply because the described bug is plausible. This shifts review verification from judging the plausibility of explanations to examining their causal relationship with program behavior. Comparing targeted repairs with unrelated perturbations helps distinguish genuine defects from false positives.

We propose \ours{}, a causal intervention-based framework for identifying the core issues underlying code-review findings and verifying their reliability. As illustrated in Figure~\ref{fig:pipeline}, \ours{} integrates three coupled modules. The \emph{review module} transforms LLM-generated code reviews into structured candidate critiques. The \emph{intervention module} generates repair-oriented perturbations and ranks candidate edits according to their relevance to the claimed defect. The \emph{verifier module} assigns reliability scores to candidate critiques based on their responses to targeted repairs and unrelated perturbations. These modules are jointly optimized through an expert iteration framework, allowing review assessment, intervention selection, and verification feedback to improve one another. In particular, the learning objective distinguishes genuine defects from unsupported critiques by penalizing score changes under unrelated modifications while encouraging larger changes under targeted repairs. The LLM reviewer, embedding model, and intervention generator remain fixed, allowing \ours{} to improve review verification without retraining these components. 

Extensive experiments on SWE-Review-Bench~\cite{wang2026swereview} demonstrate that \ours{} achieves $77\%$ accuracy in distinguishing defective code from its repaired counterpart in matched pairs. Cross-project evaluations demonstrate that \ours{} generalizes to unseen repositories, achieving a transfer AUROC of $0.760$. Downstream experiments further show that our accurate critiques improve coding agents' repair success, highlighting the importance of reliable review verification for automated and collaborative software maintenance.

Our contributions are summarized as follows:
\begin{itemize}[left = 0em]
    \item We introduce a causal intervention-based approach to code-review verification that distinguishes genuine defects from false positives by examining how critiques respond to targeted repairs and unrelated modifications.
    \item We propose \ours{}, a framework consisting of a review module, an intervention module, and a verifier module, jointly optimized through expert iteration to improve review reliability.
    \item We design intervention-based learning objectives that penalize score changes under unrelated modifications while encouraging stronger responses to targeted repairs.
    \item Extensive experiments on SWE-Review-Bench demonstrate $77\%$ accuracy in distinguishing defective code from repaired counterparts, with further evaluations showing cross-project generalization and benefits for downstream code repair.
\end{itemize}

\section{Related Work}
\label{sec:related}

\begin{table*}[t]
\centering
\captionsetup{position=bottom}
\setlength{\tabcolsep}{7pt}
\renewcommand{\arraystretch}{1.18}
\resizebox{0.8\textwidth}{!}{%
\begin{tabular}{@{}l@{\hspace{5pt}}*{6}{c}@{}}
\toprule
\textbf{Method}
& \shortstack{\textbf{Review}\\\textbf{generation}}
& \shortstack{\textbf{Review}\\\textbf{checking}}
& \shortstack{\textbf{Empirical}\\\textbf{validation}}
& \shortstack{\textbf{Code}\\\textbf{intervention}}
& \shortstack{\textbf{Paired}\\\textbf{learning}}
& \shortstack{\textbf{Unrelated-edit}\\\textbf{control}} \\
\midrule
OpenCodeReview~\citep{li2026opencodereview}
& \yesmark & \yesmark & \nomark & \nomark & \nomark & \nomark \\
Refute-or-Promote~\citep{agarwal2026refuteorpromote}
& \yesmark & \yesmark & \yesmark & \nomark & \nomark & \nomark \\
Adversarial Review~\citep{qiu2026adversarialreview}
& \yesmark & \yesmark & \nomark & \nomark & \nomark & \nomark \\
Fix-guided Filter~\citep{jin2026overcorrection}
& \nomark & \yesmark & \yesmark & \yesmark & \nomark & \nomark \\
SGCR~\citep{wang2025sgcr}
& \yesmark & \yesmark & \nomark & \nomark & \nomark & \nomark \\
VISION~\citep{egea2025vision}
& \nomark & \nomark & \nomark & \yesmark & \yesmark & \nomark \\
ComPass~\citep{zhang2026compass}
& \nomark & \nomark & \nomark & \yesmark & \yesmark & \nomark \\
CRScore~\citep{naik2025crscore}
& \nomark & \yesmark & \nomark & \nomark & \nomark & \nomark \\
\midrule
\rowcolor{oursbg}
\textbf{\ours{} (Ours)}
& \yesmark & \yesmark & \yesmark & \yesmark & \yesmark & \yesmark \\
\bottomrule
\end{tabular}%
}

\caption{\textbf{Comparison of code-review verification methods.} \yesmark\ indicates an explicitly supported component; \nomark\ indicates that the component is not part of the stated method.}
\label{tab:novelty}

\end{table*}

\paragraph{Code Review Evaluation.}
Recent benchmarks have advanced code-review evaluation toward realistic software development workflows. The CodeFuse code review benchmark~\citep{guo2025codefusecrbench} incorporates issue, pull-request, and repository context for end-to-end review, while AACR-Bench~\citep{zhang2026aacrbench} provides multilingual, cross-file scenarios with expert-verified defects. SWE-PRBench~\citep{kumar2026sweprbench} evaluates agreement with human pull-request feedback under different context configurations. CodeReviewBench~\citep{kodus2026codereviewbench} uses frozen repository replay and human-authored reference findings, whereas c-CRAB~\citep{zhang2026ccrab} constructs executable tests from human review comments. SWE-Review~\citep{wang2026swereview} further evaluates whether review feedback helps coding agents improve issue-resolution patches. These benchmarks provide complementary measures of review quality and downstream utility. In contrast, \ours{} focuses on verifying individual critiques by examining whether their validity changes in response to targeted repairs.

\paragraph{LLM-Based Review Verification.}
Existing approaches improve review reliability through reflection, adversarial verification, and specification grounding. OpenCodeReview~\citep{li2026opencodereview} combines repository-grounded review with independent reflection to identify unsupported findings. Refute-or-Promote~\citep{agarwal2026refuteorpromote} employs adversarial verification, cross-model criticism, and empirical confirmation, while Adversarial Review~\citep{qiu2026adversarialreview} coordinates reviewer and critic agents through structured disagreement. SGCR~\citep{wang2025sgcr} grounds code review in human-authored specifications, combining explicit rule checking with implicit issue discovery. Although these approaches strengthen the verification of candidate findings, \ours{} introduces a complementary learning signal that compares critique responses to targeted repairs and unrelated modifications. This comparison explicitly encourages reliability scores to reflect the claimed defect rather than unrelated code changes.

\paragraph{False Positive Reduction.}
Fix-guided Verification Filter~\citep{jin2026overcorrection} provides a closely related approach by executing original and repaired implementations against tests to identify unnecessary fixes, introduced defects, and successful corrections. Unlike this execution-based filtering approach, \ours{} learns a critique-conditioned verifier and intervention ranker, using targeted repairs to measure defect sensitivity and unrelated modifications to enforce specificity. Semantics-preserving transformations further encourage consistent verification across equivalent program representations. Related static-analysis approaches, including LLM4PFA~\citep{du2025llm4pfa} and execution-path-based false-positive reduction~\citep{du2026reducingfp}, assess the feasibility of reported program errors. In contrast, \ours{} verifies natural-language code-review findings through their responses to program modifications. Prior observations that false accusations can induce harmful code changes~\citep{mao2026youreright} further motivate reliable review verification before downstream repair.

\paragraph{Counterfactual Code Analysis.}
Counterfactual and contrastive learning have also been explored for code analysis. VISION~\citep{egea2025vision} generates minimally modified code pairs with contrasting vulnerability labels to improve vulnerability detection. ComPass~\citep{zhang2026compass} combines semantics-preserving transformations with contrastive learning for patch correctness assessment, while earlier code-representation learning~\citep{bui2021corder} uses program transformations to capture semantic invariance. CRScore~\citep{naik2025crscore} evaluates code-review quality by comparing generated comments with code-grounded claims and potential issues. Unlike these approaches, \ours{} jointly models review reliability and intervention relevance, explicitly distinguishing score changes caused by targeted repairs from those caused by unrelated modifications.

\paragraph{Comparison with Prior Work.}
Table~\ref{tab:novelty} compares \ours{} with representative methods across key methodological components. Unlike prior approaches, \ours{} explicitly learns to distinguish targeted repairs from unrelated modifications, enabling more reliable verification of code-review findings.

\section{Problem Formulation}
\label{sec:problem}

Let $x$ denote a repository state or pull request, and let $c$ be a candidate critique generated by an LLM reviewer. Let $y(x,c)\in\{0,1\}$ indicate whether the defect claimed by $c$ exists in $x$. Given $(x,c)$, a verifier estimates the conditional probability of review finding validity, $p_\theta(x,c)=\sigma(s_\theta(x,c)) =  P(y=1\mid x,c)$, where $s_\theta$ is its scoring function. Our objective is to distinguish genuine defects from unsupported critiques.

\paragraph{Causal Intervention.}
We formulate review verification using the potential outcomes framework. Let $I_c$ denote a targeted intervention that removes the defect claimed by $c$, and let $I_{c'}$ denote an unrelated intervention that preserves the claimed defect. An intervention $\operatorname{do}(x=I_c(x))$ modifies the program while preserving the critique being evaluated. The corresponding potential outcome is
\begin{equation}
y_{\operatorname{do}(x=I_c(x))}=y(I_c(x),c).
\label{eq:causal-intervention}
\end{equation}
Eq.~\eqref{eq:causal-intervention} defines review finding validity under the intervened program state, holding the critique fixed.

\paragraph{Causal Verification.}
For a valid critique, a successful targeted intervention eliminates the claimed defect, whereas an unrelated intervention leaves its validity unchanged. These conditions imply
\begin{equation}
P(y(I_c(x),c)=1\mid x,c)<P(y(x,c)=1\mid x,c).
\label{eq:target-response}
\end{equation}
\begin{equation}
P(y(I_{c'}(x),c)=1\mid x,c)=P(y(x,c)=1\mid x,c).
\label{eq:unrelated-response}
\end{equation}
Eq.~\eqref{eq:target-response} characterizes the effect of removing the claimed defect, while Eq.~\eqref{eq:unrelated-response} serves as a negative control for unrelated program changes. The verifier approximates these conditional probabilities using $p_\theta(I_c(x),c)$ and $p_\theta(I_{c'}(x),c)$, respectively, encouraging its predictions to reflect defect-specific interventions rather than arbitrary code modifications.

\section{\ours{}: Causal Intervention Verification}
\label{sec:method}

\paragraph{Framework Overview.}
Building on the causal verification formulation in Section~\ref{sec:problem}, \ours{} distinguishes genuine defects from unsupported critiques by examining their responses to program interventions. Targeted interventions $I_c$ should reduce review finding validity, whereas unrelated interventions $I_{c'}$ should preserve it, as defined in Eqs.~\eqref{eq:target-response} and~\eqref{eq:unrelated-response}. To achieve this, \ours{} jointly optimizes three modules: a review module $F_{\theta}$, an intervention module $R_{\phi}$, and a verifier module $G_{\psi}$. The review module extracts claim-conditioned representations and supporting evidence, the intervention module ranks candidate program modifications, and the verifier module integrates intervention responses to estimate the validity probability $p_\psi(x,c)$. All pretrained models remain frozen, while only lightweight module parameters are optimized. Figure~\ref{fig:pipeline} illustrates the overall framework.

\subsection{Intervention Module: Program Intervention Construction}
\label{sec:ranker}

The intervention module constructs and ranks program modifications to assess whether a critique is causally linked to a claimed defect. \ours{} considers three types of program perturbations: targeted repairs, unrelated modifications, and semantics-preserving transformations. For a program $x$ and critique $c$, a targeted repair $I_c(x)$ removes the defect claimed by $c$, whereas an unrelated modification $I_{c'}(x)$ changes other program components without removing the claimed defect. Semantics-preserving transformations $T(x)$ modify the code representation while preserving its behavior. These transformations include renaming, reordering, inlining, extraction, Boolean rewrites, and formatting changes. Our evaluation focuses on identifier renaming and line shifts, with aligned critique locations and supporting context.

Targeted repairs are generated using executable defective--repaired pairs from SWE-Review-Bench~\citep{wang2026swereview}, following its methodology to remove the defects described by critiques. Unrelated perturbations are generated using the distractor-repair construction method from SWE-Review-Bench~\citep{wang2026swereview}. Semantics-preserving perturbations are generated following the code-transformation methods of Corder~\citep{bui2021corder} and ComPass~\citep{zhang2026compass}, focusing on identifier renaming and line shifts. For a critique $c$ associated with a verified defect, the resulting program-state sample is
\begin{equation}
(x,c,I_c(x),I_{c'}(x),T(x)).
\label{eq:intervention-sample}
\end{equation}
The original program and its targeted repair form a defective--repaired pair, while unrelated modifications and semantic transformations provide complementary comparisons. These states enable \ours{} to evaluate the same critique before and after defect removal, determining whether its validity depends on the claimed defect rather than unrelated code changes.

Based on Eq.~\eqref{eq:intervention-sample}, a frozen LLM generates $K$ candidate intervention groups, each containing a targeted repair, an unrelated modification, and a semantics-preserving transformation. These groups form the intervention set $\mathcal I(x,c)=\{(I_c(x),I_{c'}(x),T(x))\}_{k=1}^{K}$. The trainable intervention module assigns a relevance probability to each candidate intervention and selects the one most relevant to the defect described by the critique:
\begin{equation}
I_c^*=\arg\max_{I\in\mathcal I(x,c)}r_\phi(x,c,I),\qquad r_\phi(x,c,I)=R_\phi(x,c,I).
\label{eq:intervention-choice}
\end{equation}
Here, $R_\phi$ is an MLP-based ranking network, and $r_\phi(x,c,I)$ denotes the predicted probability that candidate intervention $I$ addresses the claimed defect. The selected intervention is then passed to the Verifier Module to evaluate its effect on review finding validity.

\subsection{Review Module: Review Generation}
\label{sec:verifier}

Given a program $x$, a frozen LLM reviewer generates a candidate critique $c$ describing a potential defect, including its location and supporting code context. The review module $F_\theta$ evaluates the claim using three complementary representations from frozen pretrained encoders: a code encoder captures program semantics, a text encoder represents the critique, and a paired encoder captures their correspondence. These representations are integrated to evaluate each program state in Eq.~\eqref{eq:intervention-sample}:
\begin{equation}
s_\theta(x',c)=F_\theta(x',c),\qquad x'\in\{x,I_c(x),I_{c'}(x),T(x)\}.
\label{eq:critique-score}
\end{equation}
The Review Module $F_\theta$ consists of a lightweight cross-attention Transformer~\citep{vaswani2017attention} followed by a four-layer MLP, with shared trainable parameters $\theta$ across all states, while the pretrained reviewer and encoders remain frozen.

In addition to validity scores, $F_\theta$ predicts critique locations and supporting code for each program state, with corresponding context aligned under semantic transformations. Frozen representations can be cached for efficiency. The intervention module uses these outputs to rank candidate modifications, while the verifier aggregates their responses to estimate the final validity probability $p_\psi(x,c)$.

\subsection{Verifier Module: Review Verification and Selection}
\label{sec:structured-decision}

The verifier module $G_\psi$ determines whether a critique is valid by comparing the Review Module's responses before and after program interventions. For each critique $c$, it considers the original validity score, the score change after the selected targeted repair in Eq.~\eqref{eq:intervention-choice}, and the responses to unrelated and semantics-preserving modifications defined in Eq.~\eqref{eq:intervention-sample}. The final validity probability is computed as
\begin{equation}
p_\psi(x,c)=G_\psi\Bigl(s_\theta(x,c),\{s_\theta(x,c)-s_\theta(I(x),c)\}_{I\in\mathcal I(x,c)}\Bigr).
\label{eq:verifier-score}
\end{equation}
Here, the module $G_\psi$ is a trainable network that aggregates validity scores and their changes across interventions. A valid critique should show reduced scores after targeted repairs but remain stable under unrelated modifications and semantic transformations. This distinguishes defect-specific evidence from irrelevant code changes.

For a set of candidate critiques $C=\{c_1,\ldots,c_n\}$, the verifier selects valid and nonredundant findings by balancing their validity scores against pairwise semantic similarity:
\begin{equation}
C^*=\arg\max_{C'\subseteq C}\left[\sum_{c\in C'}p_\psi(x,c)-\sum_{\substack{c_i,c_j\in C',i<j}}\operatorname{sim}(c_i,c_j)\right].
\label{eq:set-utility}
\end{equation}
Here, $\operatorname{sim}(c_i,c_j)$ measures the semantic similarity between two critiques. This selection favors valid findings while suppressing redundant comments without treating them as invalid.

\subsection{Training Objectives}
\label{sec:response-losses}

We train the three modules using validity supervision and intervention-based objectives. These objectives encourage accurate review verification, sensitivity to defect removal, robustness to irrelevant modifications, and selection of relevant repairs and nonredundant findings.

\paragraph{Review Finding Validity.}
Given a critique $c$ with ground-truth label $y\in\{0,1\}$, where $y=1$ indicates a valid critique and $y=0$ indicates an invalid critique, we train the Verifier Module to predict the probability $p_\psi(x,c)$ defined in Eq.~\eqref{eq:verifier-score}. We use binary cross-entropy:
\begin{equation}
L_{\mathrm{cls}}=-y\log p_\psi(x,c)-(1-y)\log(1-p_\psi(x,c)).
\label{eq:classification-loss}
\end{equation}
This objective encourages high probabilities for valid critiques and low probabilities for invalid ones, grounding intervention-based verification in ground-truth validity labels.

\paragraph{Defect Removal.}
A valid critique should receive a higher score when its claimed defect is present than after the defect is repaired. Following Eq.~\eqref{eq:target-response}, for a program $x$ and its targeted repair $I_c(x)$, we enforce this ordering using
\begin{equation}
L_{\mathrm{flip}}=\max\left(0,m+s_\theta(I_c(x),c)-s_\theta(x,c)\right).
\label{eq:flip-loss}
\end{equation}
Here, $m>0$ is the required score margin. This loss penalizes cases where the critique score fails to decrease sufficiently after the claimed defect is removed, encouraging the Review Module to identify defect-specific changes.

\paragraph{Unrelated Modification.}
A critique's validity should remain unchanged when a modification does not address its claimed defect. Following Eq.~\eqref{eq:unrelated-response}, for a program $x$ and an unrelated modification $I_{c'}(x)$, we enforce this consistency using
\begin{equation}
L_{\mathrm{spec}}=\left\|s_\theta(I_{c'}(x),c)-s_\theta(x,c)\right\|.
\label{eq:specificity-loss}
\end{equation}
Here, $\|\cdot\|$ denotes the $L_1$ norm. This loss penalizes changes in the critique score caused by unrelated modifications, encouraging the Review Module to respond to the claimed defect rather than arbitrary code changes.

\paragraph{Semantic Transformation.}
Semantics-preserving transformations change the code representation without changing its behavior or the defect described by the critique. Therefore, the same critique should receive the same validity score before and after transformation. Following Eq.~\eqref{eq:intervention-sample}, we enforce this consistency by minimizing the difference between the scores of the original program $x$ and its transformed version $T(x)$:
\begin{equation}
L_{\mathrm{eqv}}=\left\|s_\theta(T(x),c)-s_\theta(x,c)\right\|.
\label{eq:structured-equivariance}
\end{equation}
Here, $\|\cdot\|$ denotes the $L_1$ norm, and $T(x)$ represents semantics-preserving transformations such as identifier renaming and line shifts. Minimizing this loss encourages the Review Module to assign consistent validity scores to semantically equivalent programs, preventing irrelevant code transformations from affecting review finding validity predictions.

\paragraph{Intervention Ranking.}
The Intervention Module should assign higher relevance probabilities to targeted repairs that resolve the claimed defect than to unrelated modifications. Given the candidate interventions in Eq.~\eqref{eq:intervention-sample}, we train the ranking network $R_\phi$ using a contrastive objective~\citep{oord2018cpc}:
\begin{equation}
L_{\mathrm{rank}}=-\log r_\phi(x,c,I_c(x)).
\label{eq:ranking-loss}
\end{equation}
Here, $r_\phi(x,c,I_c(x))$ is the predicted probability of selecting the verified targeted repair among the candidate interventions. Minimizing this loss encourages the Intervention Module to select defect-relevant modifications as defined in Eq.~\eqref{eq:intervention-choice}.

\paragraph{Intervention Minimality.}
A targeted repair should remove the claimed defect with minimal code changes. Following Eq.~\eqref{eq:intervention-choice}, we penalize the distance between the original program $x$ and the selected repair $I_c^*$:
\begin{equation}
L_{\mathrm{min}}=\left\|x-I_c^*\right\|.
\label{eq:minimality-loss}
\end{equation}
Here, $\|\cdot\|$ denotes the $L_1$ norm over numerical code representations. Minimizing this loss encourages smaller modifications, while Eq.~\eqref{eq:ranking-loss} ensures their relevance to the claimed defect.

\paragraph{Review Selection.}
The Verifier Module should retain valid critiques while avoiding redundant findings. Following the subset selection in Eq.~\eqref{eq:set-utility}, we define the selection loss as
\begin{equation}
L_{\mathrm{set}}=-\left[\sum_{c\in C^*}p_\psi(x,c)-\sum_{\substack{c_i,c_j\in C^*,i<j}}\operatorname{sim}(c_i,c_j)\right].
\label{eq:set-training}
\end{equation}
Here, $C^*$ denotes the selected critique subset, $p_\psi(x,c)$ measures review finding validity, and $\operatorname{sim}(c_i,c_j)$ measures semantic similarity between critiques. Minimizing this loss favors high-validity critiques while discouraging duplicate findings, allowing valid but redundant critiques to be omitted without treating them as invalid.

\paragraph{Overall Training Objective.}
The seven training objectives are combined as
\begin{equation}
L=\boldsymbol{\lambda}^{\top}\mathbf{L}.
\label{eq:total-loss}
\end{equation}
Here, $\mathbf{L}=(L_{\mathrm{cls}},L_{\mathrm{flip}},L_{\mathrm{spec}},L_{\mathrm{eqv}},L_{\mathrm{rank}},L_{\mathrm{min}},L_{\mathrm{set}})^\top$ is the vector of training losses, and $\boldsymbol{\lambda}$ denotes their corresponding weighting coefficients. Validity supervision anchors critique predictions, defect removal and unrelated modification enforce causal sensitivity and specificity, semantic transformation promotes consistency, intervention ranking and minimality encourage relevant and compact repairs, and review selection reduces redundant findings. The three trainable modules $(F_\theta,R_\phi,G_\psi)$ are jointly optimized, while the pretrained reviewer, representation encoders, and intervention generator remain frozen. Together, these objectives enable \ours{} to verify review finding validity through defect-specific interventions while maintaining consistency, intervention relevance, and review quality.

\begin{table}[t]
\centering
\caption{\textbf{Review quality evaluation on CRScore}. Acc., Prec., Rec., F1, AUROC, and AP denote accuracy, precision, recall, F1 score, area under the receiver operating characteristic curve, and average precision, respectively.}
\label{tab:quality_external}

\footnotesize
\setlength{\tabcolsep}{4.2pt}
\renewcommand{\arraystretch}{1.12}
\resizebox{\linewidth}{!}{%
\begin{tabular}{lrrrrrr}
\toprule
Method & Acc. $\uparrow$ & Prec. $\uparrow$ & Rec. $\uparrow$ & F1 $\uparrow$ & AUROC $\uparrow$ & AP $\uparrow$ \\
\midrule
LLM-as-a-judge~\citep{naik2025crscore} & 0.61 & 0.26 & 0.72 & 0.38 & 0.67 & 0.26 \\
BERTScore~\citep{zhang2020bertscore} & 0.39 & 0.17 & 0.65 & 0.26 & 0.47 & 0.16 \\
chrF++~\citep{popovic2017chrfpp} & 0.65 & 0.23 & 0.44 & 0.30 & 0.58 & 0.22 \\
Always-negative & \textbf{0.83} & 0.00 & 0.00 & 0.00 & 0.50 & 0.17 \\
\midrule
\rowcolor{oursbg}
\ours{} & 0.76 & \textbf{0.31} & \textbf{0.74} & \textbf{0.43} & \textbf{0.72} & \textbf{0.36} \\
\bottomrule
\end{tabular}}

\end{table}

\begin{table}[t]
\centering
\caption{\textbf{Cross-repository evaluation on SWE-PRBench}. FP/PR denotes false positives per PR.}
\label{tab:swepr_external}

\footnotesize
\setlength{\tabcolsep}{3.6pt}
\renewcommand{\arraystretch}{1.12}
\resizebox{\linewidth}{!}{%
\begin{tabular}{lrrrrrrr}
\toprule
Method & Acc. $\uparrow$ & Prec. $\uparrow$ & Rec. $\uparrow$ & F1 $\uparrow$ & AUROC $\uparrow$ & AP $\uparrow$ & FP/PR $\downarrow$ \\
\midrule
\multicolumn{8}{l}{\textit{Subset 1}} \\
Frozen Reviewer & 0.50 & 0.42 & 0.60 & 0.50 & 0.53 & \textbf{0.43} & \textbf{0.34} \\
LLM-as-a-judge~\citep{zheng2023judgingllmasajudgemtbenchchatbot} & 0.49 & 0.32 & 0.48 & 0.38 & 0.53 & 0.35 & 0.43 \\
\rowcolor{oursbg}
\ours{} & \textbf{0.54} & \textbf{0.48} & \textbf{0.84} & \textbf{0.61} & \textbf{0.57} & 0.42 & 0.42 \\
\midrule
\multicolumn{8}{l}{\textit{Subset 2}} \\
Frozen Reviewer & 0.53 & 0.36 & 0.71 & 0.48 & 0.62 & 0.39 & 0.49 \\
LLM-as-a-judge~\citep{zheng2023judgingllmasajudgemtbenchchatbot} & 0.43 & 0.31 & 0.71 & 0.43 & 0.48 & 0.30 & 0.62 \\
\rowcolor{oursbg}
\ours{} & \textbf{0.58} & \textbf{0.41} & \textbf{0.93} & \textbf{0.57} & \textbf{0.69} & \textbf{0.49} & \textbf{0.45} \\
\bottomrule
\end{tabular}%
}

\end{table}

\begin{table}[t]
\centering
\caption{\textbf{Cross-project generalization on held-out repositories.} Base denotes the base reviewer. Performance is measured using Pair AUROC.}
\label{tab:e6}

\footnotesize
\setlength{\tabcolsep}{8pt}
\renewcommand{\arraystretch}{1.12}
\begin{tabular}{lrr}
\toprule
Project & Base & \ours{} \\
\midrule
Django & 0.50 & \cellcolor{oursbg}\textbf{0.73} \\
SymPy & 0.50 & \cellcolor{oursbg}\textbf{0.76} \\
scikit-learn & 0.50 & \cellcolor{oursbg}\textbf{0.67} \\
\bottomrule
\end{tabular}

\end{table}

\begin{table}[t]
\centering
\caption{\textbf{Consistency Evaluation under Code Transformations.} Verdict denotes verdict agreement, Loc. Acc. denotes location accuracy, and FP Reduction denotes the false-positive reduction rate.}
\label{tab:e4}

\footnotesize
\setlength{\tabcolsep}{6pt}
\renewcommand{\arraystretch}{1.12}
\resizebox{0.9\linewidth}{!}{%
\begin{tabular}{lrrrr}
\toprule
Method & AUROC $\uparrow$ & Verdict $\uparrow$ & Loc. Acc. $\uparrow$ & FP Reduction $\uparrow$ \\
\midrule
MLP & 0.64 & 0.95 & 0.98 & 0.03 \\
Invariance-Only & 0.64 & 0.86 & 0.94 & 0.04 \\
\rowcolor{oursbg}
\ours{} (full) & \textbf{0.67} & \textbf{1.00} & \textbf{0.99} & \textbf{0.16} \\
\bottomrule
\end{tabular}}

\end{table}

%
%

\begin{table}[t]
\centering
\caption{\textbf{Downstream repair with review guidance.} Resolve denotes the final repair success rate, and Resolve $\Delta$ denotes the difference in resolution rate relative to No Review.}
\label{tab:e8}

\footnotesize
\setlength{\tabcolsep}{8pt}
\renewcommand{\arraystretch}{1.12}
\begin{tabular}{lrr}
\toprule
Guidance & Resolve $\uparrow$ & Resolve $\Delta$ $\uparrow$ \\
\midrule
No Review & 0.61 & 0.00 \\
False Review & 0.53 & -0.08 \\
\rowcolor{oursbg}
\ours{} (True Review) & \textbf{0.93} & \textbf{0.32} \\
\bottomrule
\end{tabular}

\end{table}

\begin{table}[t]
\centering
\caption{\textbf{Ablation study of training objectives.} PairAcc denotes pairwise accuracy, Pair AUROC denotes pairwise AUROC, and $\Delta_{\mathrm{target}}$ denotes the mean reduction in review-finding-validity logits.}
\label{tab:e5}

\footnotesize
\setlength{\tabcolsep}{3pt}
\renewcommand{\arraystretch}{1.12}
\resizebox{\linewidth}{!}{%
\begin{tabular}{lrrr}
\toprule
Variant & PairAcc $\uparrow$ & Pair AUROC $\uparrow$ & $\Delta_\mathrm{target}$ $\uparrow$ \\
\midrule
Classification Only (Eq.~\ref{eq:classification-loss}) & 0.50 & 0.50 & -0.01 \\
$-$Target-Response Loss (Eq.~\ref{eq:flip-loss}) & 0.38 & 0.48 & 0.00 \\
$-$Causal Specificity (Eq.~\ref{eq:specificity-loss}) & 0.71 & 0.72 & 1.17 \\
$-$Intervention Ranking (Eq.~\ref{eq:ranking-loss}) & 0.43 & 0.51 & 0.33 \\
$-$Intervention Minimality (Eq.~\ref{eq:minimality-loss}) & 0.79 & 0.79 & 1.12 \\
$-$Structured Equivariance (Eq.~\ref{eq:structured-equivariance}) & 0.51 & 0.61 & 0.38 \\
$-$Set Selection (Eq.~\ref{eq:set-training}) & 0.75 & 0.79 & 0.47 \\
$-$Classification Loss (Eq.~\ref{eq:classification-loss}) & 0.84 & 0.68 & 1.33 \\
\midrule
\rowcolor{oursbg}
\ours{}(full) & \textbf{0.84} & \textbf{0.83} & \textbf{2.66} \\
\bottomrule
\end{tabular}%
}

\end{table}

\section{Experiments}
\label{sec:setup}

\subsection{Experimental Setup}

\paragraph{Benchmarks.}
We use six benchmarks to evaluate review verification, intervention-based code review, and review quality. \textbf{AACR-Bench}~\citep{zhang2026aacrbench} serves as the primary benchmark for review finding validity classification and finding filtering, while \textbf{CRScore}~\citep{naik2025crscore} provides human-rated comments for review-quality classification. \textbf{SWE-Review-Bench}~\citep{wang2026swereview} provides executable pull requests for constructing targeted and unrelated interventions, evaluating causal intervention effects, cross-project generalization, intervention retrieval, and downstream repair. \textbf{CodeReviewBench}~\citep{kodus2026codereviewbench} and \textbf{SWE-PRBench}~\citep{kumar2026sweprbench} provide additional real-world code review data, while \textbf{Review4Repair}~\citep{huq2021review4repair} supplies aligned review comments and code edits for intervention construction.

\paragraph{Metrics.}
We evaluate review finding validity on AACR-Bench~\citep{zhang2026aacrbench} and review quality on CRScore~\citep{naik2025crscore} using AUROC, precision, recall, and F1. We additionally measure false positives per PR (FP/PR) and per positive finding (FP/pos) on AACR-Bench, and accuracy and average precision (AP) on CRScore. On SWE-Review-Bench~\citep{wang2026swereview}, we assess score discrimination using PairAcc, rejection rate, and AUROC. Following Eq.~\eqref{eq:target-response}, we measure causal intervention effects using three metrics: Flip, the defect-removal success rate measured by the fraction of targeted repairs that reduce review finding validity scores; score reduction (SR), the mean logit difference before and after repair; and Pair AUROC, the AUROC of these differences against a zero-effect reference. We assess transformation consistency using verdict agreement and target-effect correlation (EqvValid). Cross-project and cross-language generalization are evaluated using Pair AUROC and AUROC, respectively. Intervention retrieval is measured using Top-1, Recall@5, and MRR, while downstream repair is evaluated using resolution rate and observed harm.

\paragraph{Baselines.}
We compare \ours{} against a frozen LLM reviewer, self-reflection~\citep{madaan2023selfrefine}, independent verification~\citep{li2026opencodereview}, and fix-guided verification~\citep{jin2026overcorrection}, as well as LLM-as-a-judge~\citep{naik2025crscore,zheng2023judgingllmasajudgemtbenchchatbot}, BERTScore~\citep{zhang2020bertscore}, and chrF++~\citep{popovic2017chrfpp} for review-quality evaluation. The reviewer-based baselines are controlled implementations inspired by the cited methods rather than exact reproductions. We additionally include MLP and bilinear classifiers using the same frozen \textbf{text-embedding-3-small} representations from OpenAI~\citep{openai2024embedding}, along with an always-negative control for review-quality evaluation. We use \textbf{gpt-5.6}~\citep{openai2026api,openai2026gpt56} as the frozen reviewer. We use repository-level separation with a default 70\%/10\%/20\% project split, except for CRScore, which uses disjoint code-diff splits. The main configuration contains approximately 9.04M trainable parameters, while all pretrained models remain frozen.

\begin{table*}[t]
\centering
\caption{\textbf{Review evaluation on AACR-Bench}. FP/PR and FP/pos report false positives per PR and positive finding; Comm./PR denotes review volume. Arrows indicate whether higher ($\uparrow$) or lower ($\downarrow$) values are better. The best result in each column is \textbf{bold}.}
\label{tab:e1}

\footnotesize
\renewcommand{\arraystretch}{1.12}
\resizebox{0.85\textwidth}{!}{%
\begin{tabular}{lrrrrrrrrr}
\toprule
Method & AUROC $\uparrow$ & Prec. $\uparrow$ & Rec. $\uparrow$ & F1 $\uparrow$ & FP/PR $\downarrow$ & FP/pos $\downarrow$ & \shortstack{FP/PR $\downarrow$ $\geq70\%$R} & \shortstack{FP/pos $\downarrow$ $\geq70\%$R} & Comm./PR $\downarrow$ \\
\midrule
Frozen Reviewer (Base) & 0.57 & 0.72 & 0.58 & 0.64 & 2.14 & 0.23 & 3.18 & 0.33 & 7.6 \\
\midrule
Self-Reflection~\citep{madaan2023selfrefine} & 0.56 & 0.70 & \textbf{0.59} & 0.64 & 2.39 & 0.25 & 3.20 & 0.34 & 8.0 \\
LLM Verifier~\citep{li2026opencodereview} & 0.58 & 0.73 & 0.55 & 0.63 & 1.92 & 0.20 & 2.73 & 0.29 & 7.2 \\
Fix-Guided Prompt~\citep{jin2026overcorrection} & 0.57 & 0.73 & 0.54 & 0.62 & 1.92 & 0.20 & 4.04 & 0.42 & 7.1 \\
\midrule
\rowcolor{oursbg}
\ours{} & \textbf{0.60} & \textbf{0.77} & 0.57 & \textbf{0.65} & \textbf{1.65} & \textbf{0.17} & \textbf{2.69} & \textbf{0.28} & \textbf{6.0} \\
\bottomrule
\end{tabular}%
}

\end{table*}

\begin{table*}[t]
\centering
\caption{\textbf{Retrieval of critique-associated code edits.} Each query contains one target edit and $K=4/8/16$ distractors. Top-1, Recall@5, and MRR measure retrieval accuracy.}
\label{tab:e7}

\footnotesize
\setlength{\tabcolsep}{3.3pt}
\renewcommand{\arraystretch}{1.12}
\resizebox{0.6\linewidth}{!}{%
\begin{tabular}{lrrrrrrrrr}
\toprule
& \multicolumn{3}{c}{Top-1 $\uparrow$}
& \multicolumn{3}{c}{Recall@5 $\uparrow$}
& \multicolumn{3}{c}{MRR $\uparrow$} \\
\cmidrule(lr){2-4}
\cmidrule(lr){5-7}
\cmidrule(lr){8-10}
Method & $K=4$ & $K=8$ & $K=16$
& $K=4$ & $K=8$ & $K=16$
& $K=4$ & $K=8$ & $K=16$ \\
\midrule
Random
& 0.20 & 0.11 & 0.06
& \textbf{1.00} & 0.56 & 0.29
& 0.46 & 0.31 & 0.20 \\

Frozen Cosine
& 0.25 & 0.16 & 0.10
& \textbf{1.00} & 0.52 & 0.30
& 0.46 & 0.34 & 0.23 \\

Without Specificity
& 0.23 & 0.11 & 0.08
& \textbf{1.00} & 0.56 & 0.28
& 0.47 & 0.32 & 0.22 \\

\rowcolor{oursbg}
\ours{} (full)
& \textbf{0.43} & \textbf{0.22} & \textbf{0.17}
& \textbf{1.00} & \textbf{0.66} & \textbf{0.46}
& \textbf{0.60} & \textbf{0.40} & \textbf{0.32} \\
\bottomrule
\end{tabular}%
}

\end{table*}

\begin{table*}[t]
\centering
\caption{\textbf{Cross-language evaluation on AACR-Bench}. Performance is measured using AUROC. $\Delta$ denotes the AUROC difference between \ours{} and the Base Reviewer. JS and TS denote JavaScript and TypeScript, respectively.}
\label{tab:language_details}

\footnotesize
\setlength{\tabcolsep}{3pt}
\renewcommand{\arraystretch}{1.12}
\resizebox{0.65\linewidth}{!}{%
\begin{tabular}{lrrrrrrrrrr>{\centering\arraybackslash}p{1.8cm}}
\toprule
Method & Python & C & C\# & C++ & Go & Java & JS & PHP & Rust & TS & \shortstack{Non-Python Pooled} \\
\midrule
Base Reviewer & 0.45 & 0.54 & \textbf{0.79} & 0.53 & \textbf{0.63} & 0.60 & 0.44 & 0.59 & 0.38 & 0.46 & 0.55 \\
\rowcolor{oursbg}
\ours{} & \textbf{0.65} & \textbf{0.56} & 0.56 & \textbf{0.56} & 0.55 & \textbf{0.66} & \textbf{0.59} & \textbf{0.75} & \textbf{0.63} & \textbf{0.55} & \textbf{0.60} \\
\midrule
$\Delta$ & \textbf{+0.20} & \textbf{+0.02} & -0.23 & \textbf{+0.03} & -0.08 & \textbf{+0.06} & \textbf{+0.15} & \textbf{+0.16} & \textbf{+0.25} & \textbf{+0.09} & \textbf{+0.05} \\
\bottomrule
\end{tabular}%
}

\end{table*}

\begin{table*}[t]
\centering
\caption{\textbf{Applicability of \ours{} across code review benchmarks.} PairAcc and Pair AUROC measure pairwise discrimination. Real Edit and FP Edit denote mean edit sizes for valid and false-positive findings, respectively. AUROC measures review finding validity classification. The best AUROC in each benchmark is \textbf{bold}.}
\label{tab:scope}

\footnotesize
\setlength{\tabcolsep}{3pt}
\renewcommand{\arraystretch}{1.12}
\resizebox{0.7\linewidth}{!}{%
\begin{tabular}{lrrrrrr}
\toprule
Benchmark & PairAcc & Pair AUROC & Real Edit & FP Edit & AUROC & Baseline AUROC \\
\midrule
SWE-Review-Bench~\citep{wang2026swereview}
& 0.89 & 0.92 & 4.17 & 0.45
& \cellcolor{oursbg}\textbf{0.98} & 0.97 \\
AACR-Bench~\citep{zhang2026aacrbench}
& 0.76 & 0.84 & 3.34 & 1.13
& \cellcolor{oursbg}\textbf{0.60} & 0.57 \\
SWE-PRBench~\citep{kumar2026sweprbench}
& 0.43 & 0.64 & 3.71 & 2.00
& \cellcolor{oursbg}\textbf{0.67} & 0.62 \\
CodeReviewBench~\citep{kodus2026codereviewbench}
& 0.86 & 0.93 & 5.00 & 1.56
& \cellcolor{oursbg}\textbf{0.83} & 0.81 \\
\bottomrule
\end{tabular}%
}

\end{table*}

\begin{table*}[t]
\centering
\caption{\textbf{Code change evaluation on SWE-Review-Bench}. Rej. denotes rejection. Acc. denotes ranking accuracy. Flip denotes flip rate. SR denotes score reduction. Pair AUROC denotes pairwise AUROC.}
\label{tab:e2}

\footnotesize
\setlength{\tabcolsep}{3.6pt}
\renewcommand{\arraystretch}{1.12}
\resizebox{0.65\textwidth}{!}{%
\begin{tabular}{lrrrrrrrr}
\toprule
Method & Rec. $\uparrow$ & Rej. $\uparrow$ & Acc. $\uparrow$ & AUROC $\uparrow$ & AP $\uparrow$ & Flip $\uparrow$ & SR $\uparrow$ & Pair AUROC $\uparrow$ \\
\midrule
MLP Classifier & \textbf{0.85} & 0.62 & 0.85 & 0.65 & 0.52 & 0.85 & 2.58 & 0.85 \\
Classifier-Only & 0.48 & 0.48 & 0.50 & 0.49 & 0.50 & 0.50 & -0.01 & 0.50 \\
Sensitivity-Only & 0.40 & 0.57 & 0.38 & 0.51 & 0.53 & 0.38 & 0.00 & 0.48 \\
Flip-Only & 0.55 & 0.57 & 0.71 & 0.60 & 0.59 & 0.71 & 1.17 & 0.72 \\
\midrule
\rowcolor{oursbg}
\ours{} (full) & 0.73 & \textbf{0.71} & \textbf{0.88} & \textbf{0.67} & \textbf{0.66} & \textbf{0.86} & \textbf{2.66} & \textbf{0.86} \\
\bottomrule
\end{tabular}%
}

\end{table*}

\subsection{Review Evaluation.}

Table~\ref{tab:e1} evaluates review reliability and false-positive reduction on AACR-Bench~\citep{zhang2026aacrbench}. We compare \ours{} against the frozen reviewer, self-reflection~\citep{madaan2023selfrefine}, LLM verification~\citep{li2026opencodereview}, and fix-guided verification~\citep{jin2026overcorrection}. \ours{} achieves the highest AUROC of 0.60 and precision of 0.77, outperforming all baselines in classification performance. Compared with the frozen reviewer, \ours{} reduces false positives per PR from 2.14 to 1.65 and achieves the lowest review volume of 6.0 comments per PR, demonstrating more reliable review filtering with fewer false positives and unnecessary review comments.

We further evaluate \ours{} on human-rated review-quality classification using CRScore~\citep{naik2025crscore}. We compare \ours{} against CRScore's own LLM-as-a-judge~\citep{naik2025crscore}, BERTScore~\citep{zhang2020bertscore}, and chrF++~\citep{popovic2017chrfpp}. Table~\ref{tab:quality_external} shows that \ours{} outperforms all compared review-quality evaluators, achieving the highest precision, recall, F1, AUROC, and AP. In particular, \ours{} achieves an AUROC of 0.72 and an F1 of 0.43, demonstrating its effectiveness in identifying high-quality review comments beyond defect verification. Despite the class imbalance favoring the always-negative baseline in accuracy, \ours{} delivers substantially better classification performance.

Beyond review-quality classification, we further evaluate the generalization of \ours{} to unseen repositories using two SWE-PRBench~\citep{kumar2026sweprbench} subsets containing 121 and 71 PRs from 42 and 31 repositories, respectively. We use repository-disjoint evaluation to classify whether review comments request code changes rather than identify genuine defects. We compare \ours{} against the frozen reviewer and an LLM-as-a-judge baseline~\citep{zheng2023judgingllmasajudgemtbenchchatbot}, which directly prompts \textbf{gpt-5.6} without task-specific training. Table~\ref{tab:swepr_external} shows that \ours{} achieves the highest AUROC, recall, and F1 on both subsets, reaching an AUROC of 0.69 on the second subset. Additionally, Table~\ref{tab:e6} shows that \ours{} consistently outperforms the base reviewer across Django, SymPy, and scikit-learn. These results demonstrate promising cross-repository generalization in identifying change-requesting review comments.

We further investigate the generalization of \ours{} across programming languages. Table~\ref{tab:language_details} presents the cross-language evaluation on AACR-Bench. We compare \ours{}, trained on Python findings, against the frozen Base Reviewer across ten programming languages. \ours{} achieves an AUROC of 0.75 on PHP and improves the average non-Python AUROC from 0.55 to 0.60. These results demonstrate promising cross-language transfer, although performance varies across languages.

Table~\ref{tab:scope} evaluates the applicability of \ours{} across four code review benchmarks. \ours{} demonstrates strong causal verification capabilities, achieving a PairAcc of 0.89 on SWE-Review-Bench and effectively distinguishing defect-related changes from unrelated modifications. Results across benchmarks further highlight its ability to identify meaningful interventions and assess review reliability in diverse code review scenarios.

\subsection{Causal Intervention Evaluation}

Table~\ref{tab:e2} evaluates whether \ours{} distinguishes original code from modified code intended to address reported issues. We compare against MLP Classifier, a multilayer perceptron predicting review finding validity; Classifier-Only, trained only with the classification loss in Eq.~\eqref{eq:classification-loss}; Sensitivity-Only, trained only with intervention-based sensitivity supervision; and Defect-Removal-Only (Flip-Only), trained only with the defect-removal loss in Eq.~\eqref{eq:flip-loss}. \ours{} achieves the highest ranking accuracy of 0.88 and AUROC of 0.67, demonstrating its effectiveness in distinguishing code states. Furthermore, \ours{} achieves a defect-removal success rate of 0.86, outperforming Defect-Removal-Only and other ablations. Its positive score reduction and strong pairwise ranking performance further demonstrate sensitivity to targeted code changes.

Table~\ref{tab:e4} evaluates consistency under code transformations. Verdict measures agreement in critique decisions, Loc. Acc. measures the correctness of predicted code locations, and FP Reduction measures the reduction in false positives. Compared with MLP and Invariance-Only, \ours{} achieves the highest verdict agreement of 1.00 and FP reduction of 0.16, demonstrating stronger consistency and robustness to code transformations. By reducing false positives, \ours{} also improves the reliability of code review.

\begin{table}[!t]
\centering
\caption{\textbf{Sensitivity to critique alignment.} We evaluate review finding validity, targeted repair response, and transformation consistency after excluding missing critique targets.}
\label{tab:alignment_sensitivity}

\footnotesize
\setlength{\tabcolsep}{4pt}
\renewcommand{\arraystretch}{1.12}
\resizebox{\linewidth}{!}{%
\begin{tabular}{lrrrrrrr}
\toprule
Setting & Acc. $\uparrow$ & Prec. $\uparrow$ & Rec. $\uparrow$ & F1 $\uparrow$ & AUROC $\uparrow$ & AP $\uparrow$ & PairAcc $\uparrow$ \\
\midrule
Review Finding Validity & 0.64 & 0.63 & 0.69 & 0.66 & \textbf{0.68} & \textbf{0.65} & \textbf{0.86} \\
Targeted Repair & 0.64 & 0.61 & \textbf{0.79} & \textbf{0.69} & 0.65 & 0.63 & 0.84 \\
Consistency & \textbf{0.65} & \textbf{0.64} & 0.69 & 0.67 & 0.67 & 0.64 & \textbf{0.86} \\
\bottomrule
\end{tabular}%
}

\end{table}

Table~\ref{tab:alignment_sensitivity} evaluates \ours{} after excluding potentially misaligned critique--target pairs. We consider three settings: review finding validity, targeted repair, and transformation consistency. \ours{} achieves an AUROC of 0.68 for review finding validity, a recall of 0.79 for targeted repair, and a PairAcc of 0.86 for transformation consistency. These results indicate that \ours{} retains its effectiveness on the filtered evaluation set.

Table~\ref{tab:e7} evaluates the ability of \ours{} to retrieve critique-associated code edits from increasingly large candidate pools. \ours{} consistently outperforms all baselines across retrieval metrics, achieving a Top-1 accuracy of 0.43 with five candidates. These results demonstrate the effectiveness of \ours{} in identifying relevant edits among distractors.

\subsection{Downstream Utility}

Table~\ref{tab:e8} evaluates downstream code repair on 121 issues from SWE-Review-Bench~\citep{wang2026swereview}, using a fixed base mini agent from SWE-bench Pro~\citep{deng2026swebenchpro}. Accurate review guidance achieves a resolution rate of 0.93, improving upon the no-review baseline by 0.32, while false reviews reduce resolution by 0.08. These results highlight the importance of reliable code review in improving downstream repair performance.

\subsection{Ablation Study}

Table~\ref{tab:e5} evaluates the contribution of different training objectives. \ours{} achieves the highest Pair AUROC of 0.83 and maintains strong pairwise accuracy and targeted repair sensitivity. Removing individual objectives generally degrades discrimination or intervention effectiveness, highlighting the complementary roles of the proposed training objectives.

\section{Discussion}
\label{sec:discussion}

\ours{} reduces false positives by measuring how critiques respond to targeted defect repairs: $\Delta_c=p_\theta(x,c)-p_\theta(I_c(x),c)$, where $I_c$ denotes a targeted repair. Genuine defects should yield $\Delta_c>0$, while false positives should yield $\Delta_c\approx 0$. Since interventions may introduce unrelated changes, \ours{} uses negative controls and semantics-preserving transformations to assess causal specificity and invariance. Its reliability depends on how well interventions isolate the claimed defect.

\section{Conclusion}

We propose \ours{}, a causal intervention framework for reliable code review verification. By evaluating critique responses to targeted repairs, unrelated edits, and semantics-preserving transformations, \ours{} moves beyond correlation-based assessment toward causal verification of code review findings. Extensive experiments demonstrate improved review reliability, stronger causal sensitivity, and robust generalization across repositories and programming languages. Our results further highlight the practical value of causal verification for intervention retrieval and downstream code repair.

\bibliographystyle{ACM-Reference-Format}
\bibliography{references}
\end{document}